\documentclass[twocolumn,showpacs,preprintnumbers,amsmath,amssymb]{revtex4-1}

\usepackage{dcolumn}
\usepackage{bm}

\usepackage{color}
\usepackage{graphicx}
\usepackage{natbib}
\usepackage{subfigure}
\usepackage{cases}
\usepackage{epstopdf}

\usepackage{dcolumn} 
\newcommand{\beq}{\begin{equation}}
\newcommand{\eeq}{\end{equation}}

\begin{document}
\def\bfB{\mbox{\bf B}}
\def\bfQ{\mbox{\bf Q}}
\def\bfD{\mbox{\bf D}}
\def\etal{\mbox{\it et al}}

\title{Searching for the fastest dynamo: \\ Laminar ABC flows}
\author{Alexandros Alexakis}

\affiliation{Laboratoire de Physique Statistique de l'Ecole Normale
Sup\'erieure, UMR CNRS 8550, 24 Rue Lhomond, 75006 Paris Cedex 05, France.}

\date{\today}

\begin{abstract}
The growth rate of the dynamo instability as a function of the magnetic Reynolds number $R_{_M}$
is investigated by means of numerical simulations for the family of the ABC flows  and 
for 2 different forcing scales. 
For the ABC flows that are driven at the largest available 
length scale it is found that as the magnetic Reynolds number is increased:
        (a) The flow that results first in dynamo is the $2\frac{1}{2}$D flow for which  A=B and C=0 
            (and all permutations).
        (b) The second type of flow that results in dynamo is the one for which $A=B\simeq 2C/5$ 
            (and  permutations).
        (c) The most symmetric flow A=B=C is the third type of flow that results in dynamo.
        (d) As $R_{_M}$ is increased, the A=B=C flow stops being a dynamo and transitions from a local 
            maximum to a third-order saddle point.
        (e) At larger $R_{_M}$ the A=B=C flow re-establishes its self as a dynamo but remains a saddle point.
        (f) At the largest examined $R_{_M}$ the growth rate of the $2\frac{1}{2}$D flows starts to decay,
            the A=B=C flow comes close to a local maximum again and
            the flow  $A=B\simeq 2C/5$ (and permutations) results in the fastest dynamo with growth 
            rate $\gamma \simeq 0.12$ at the largest examined $R_{_M}$.
For the ABC flows that are driven at the second largest available length scale it is found that
        (a) the 2$\frac{1}{2}$D  flows A=B, C=0 (and permutations) are again the first flows that result 
            in dynamo with a decreased onset.
        (b) The most symmetric flow A=B=C is the second type of flow that results in  dynamo. It is and 
            remains a local maximum.
        (c) At larger  $Rm$ the flow $A=B\simeq 2C/5$ (and permutations) appears as the 
            third type of flow that results in dynamo. As $R_{_M}$ is increased it becomes the flow with 
            the largest growth rate.
The growth rates appear to have some correlation with the Lyaponov exponents but constructive re-folding
of the field lines appears equally important in determining the fasted dynamo flow.
\end{abstract} 
%
%
%
\maketitle
%
\section{Introduction}
%
Magnetic dynamo is the process through which an electrically conducting fluid amplifies and maintains 
magnetic energy against Ohmic dissipation  by continuously stretching and re-folding the magnetic 
field lines \cite{Zeldo1990,Moffatt1978}. This process is considered to be the main mechanism for the generation 
of  magnetic energy in the universe. It is present in the intergalactic and interstellar medium, in accretion 
disks and in the interiors of stars and planets. It has also been realized recently in different 
laboratory experiments \cite{Gailitis2001,Stieglitz2001,VKS}. 
The flows in these examples vary in structure and the generated magnetic fields exhibit a large variety of
structural and temporal behavior. It is then desirable to understand which properties of a flow are 
important for the amplification of magnetic  energy and how do they effect the dynamical behavior of the 
magnetic field. This question is of particular  interest for the dynamo experiments for which optimizing 
the flow is important for achieving dynamo at small energy injection rates \cite{Ravelet2005}. 

In theoretical studies, various flows have been examined analytically and numerically both in the laminar and 
in the turbulent regime. 
The Ponomarenko \cite{Ponomarenko1973}, the ABC \cite{Arnold1965,Beltrami1889,Childress1970}, the Roberts 
\cite{Roberts1970,Roberts1972}, the Taylor-Green \cite{TG}, 
and the Archontis flow \cite{Archontis2007} are some of the flows that have been shown to result 
successfully in dynamo action provided that the magnetic Reynolds number $R_{_M}$  (the ratio of the large-scale 
velocity time scale to the large-scale diffusivity time scale) is sufficiently large. 
The choice of flow for study was motivated either by its similarity to astrophysical flows or due to  its simplicity that 
allowed analytical treatment or made the investigation more tractable numerically. Other than this practical 
motivation there is no mathematical justification for preferring one flow over an other.

This lack of mathematical reasoning motivates this work. 
Over a family of flows of finite energy and vorticity not all members are as efficient in producing dynamo action. 
It is then expected that a flow in this family exist that is  optimal for dynamo action.
Finding and investigating the properties of such an optimal flow can then reveal which mechanisms are important for 
magnetic field amplification. 
How an optimal flow is defined 
is described in the next section where the general problem is formulated in detail.

\section{formulation}

At the  early stages of the dynamo, when the Lorentz force is too weak to act back on the flow,
the evolution of the magnetic field is given by the linear advection diffusion equation
\begin{equation}\label{dynamo}
\partial_t {\bf b} + {\bf u \cdot \nabla b} = {\bf b \cdot \nabla u}           +\eta \nabla^2 {\bf b},
\end{equation}
where ${\bf b}$ is the magnetic vector field, ${\bf u}$ is the velocity field and $\eta$ is the magnetic diffusivity.
The advection term in the left hand side of Eq.(\ref{dynamo}) is responsible for the mixing of the magnetic field lines. 
The first term in the right hand side is the stretching term that is responsible for the increase of the magnetic energy 
while the last term is responsible for the destruction of magnetic energy due to diffusion. 
Since the equation for the magnetic field is linear it expected that after some transient behavior
the amplitude of the magnetic field will grow or decay at an exponential rate
\beq\label{str_eig}
{ \bf b} \sim \tilde{\bf b}({\bf x},t) \, e^{\gamma t },
\eeq
where $\gamma$ is the growth rate and $\tilde{\bf b}$ is a bounded function in time.
For  steady velocity fields that will be examined here $\tilde{\bf b}$ is either time
independent or a periodic function of time.

The growth rate $\gamma$ and its dependence on the flow parameters is the primary interest in this work.
Given the functional shape of ${\bf u}$ the only control parameter in the system is the magnetic Reynolds number
$R_{_M}$ that in this work it is defined as
\begin{equation} 
\label{RM}
R_{_M} \equiv \frac{U }{ \eta k_u}. 
\end{equation}
$\eta$ is the magnetic diffusivity. $U$ is the amplitude of the velocity field that is defined as
\beq
U\equiv \langle {\bf u\cdot u}\rangle^{1/2}, 
\eeq
where the angular brackets stand for spatial average. 
$k_u$ is the velocity inverse length-scale 
that we define through the vorticity of the flow ${\bf w = \nabla \times u}$ as 
\beq
k_u \equiv \langle {\bf w\cdot w} \rangle^{1/2}/U .
\eeq

Exploring the dependence of $\gamma$ on $R_{_M}$ has been the subject of extensive research. 
For sufficiently small $R_{_M}$ the diffusion term in Eq.(\ref{dynamo}) will dominate and 
magnetic energy will decrease exponentially with decay rate $-\gamma \sim Uk_u R_{_M}^{-1}$
(for $R_{_M}\ll 1$).  
As $R_{_M}$ is increased the stretching term becomes important and above a critical value a flow can 
become an effective dynamo ($\gamma>0$). This value of $R_{_{M}}$ will be referred to as the critical 
magnetic Reynolds number and will be denoted as $R_{_{MC}}$. Finding the flow that minimizes $R_{_{MC}}$ 
is of importance for laboratory dynamo experiments on account of $R_{_M}$ being an increasing function of power
consumption which is an increasing function of cost.

For large values of the magnetic Reynolds number the problem becomes increasingly complex with 
the number of degrees of freedom involved increasing like $R_{_M}^{3/2}$. 
Due to this complexity there is no general analytic way to estimate the growth rate of a dynamo
(with the exception of some special cases).
Nonetheless, anti-dynamo theorems \cite{Cowling1934,Zeldovich1957} developed in the last century and upper
bounds on the growth rate \cite{Backus1958,Childress1969,Proctor1979,Proctor2004} have been proven useful in
excluding certain classes of flows from giving dynamo action or restricting the scaling of $\gamma$ with $R_{_M}$.
From the Anti-dynamo theorems two important results  that are relevant in
this work are mentioned here.

First,
          flows with only two non-zero components of the velocity field can not result in dynamo for any
          value of $R_{_M}$ \cite{Zeldovich1957}. Thus, for these flows there is no critical Reynolds number.
          In the present work we will refer to these flows as 2D flows (even if there is spatial dependence in the
          third direction).

Second,
         time independent flows for which all three components of the velocity ($u_x,u_y,u_z$) are non-zero
         but depend only on two of the spatial components (say $x,y$) can result in dynamo, but due to the
         absence of chaotic flow lines the dynamo growth rate will tend to a non-positive
         value as $R_{_M}$ tends to infinity \cite{Friedlander1991,Vishik1989}. For these flows it is thus expected that
         \beq
         \gamma  = {o}(1)\,\, Uk_u,
         \eeq
         where the symbol ``${o}(1)$" stands for ``smaller than order one". This dependence however can be a very slowly
         decreasing function of $R_{_M}$ \cite{Soward1987}. These flows are referred to as $2\frac{1}{2}D$ flows, and the resulting 
         dynamo is referred to as a slow dynamo.

         However, besides these classes of flows typical three-dimensional flows with a complex streamline
         topology are expected to be dynamos at infinitely large $R_{_M}$ \cite{Finn1988}. 
         For such flows the growth rate will approach a value that depends only on the amplitude, length-scale 
         and structure of the velocity field and not on the magnetic diffusion $\eta$ {\it ie }:
         \beq
         \gamma  = \mathcal{O}(1)\,\, Uk_u.
         \eeq
         where the symbol ``$\mathcal{O}(1)$" stands for ``same order as one". 
         Such flows for which the dynamo growth rate tends to a positive value as $R_{_M}$ tends to infinity 
         will be called fast dynamo flows.

With these restrictions in mind a definition of ``an optimal flow" can be given.
The choice of optimization will of course depends on the application in mind.
For example, an optimal flow can be based on $R_{_{MC}}$ or on $\gamma$ leading to different answers. 
Here we will restrict to the following questions:
Given a family of flows of fixed velocity amplitude $U$ and length-scale $k_u$ 
            (i)    which member has the smallest critical magnetic Reynolds number $R_{_{MC}}$, 
            (ii)   given $R_{_M}$ which member has the largest growth rate $\gamma/(Uk_u)$, 
and finally (iii)  which flow leads to the largest growth rate in the limit $R_{_M}\to\infty$.
As it is shown later, these questions do not have the same answer.

Finally we need to restrict the family of flows that are going to be investigated.
Since the estimate of the growth rate $\gamma$ and $R_{_{MC}}$ needs to be done numerically
addressing the questions above for a large family of flows is formidable even for present day computing. 
It is then preferable to restrict to smaller families that are however good candidates for fast dynamo action 
based on their properties. 

For a fast dynamo the role of chaoticity and helicity of the flow has been emphasized as important ingredients.
Chaoticity, the exponential stretching of fluid elements, is a necessary ingredient a for fast dynamo
\cite{Vishik1989,Friedlander1991}. However, it is not sufficient. Time dependent 2D flows can result in chaos
(positive Lyaponov exponent) but can be excluded from  dynamo flows based on the first anti-dynamo theorem
mentioned here. The reason for this behavior is that the flow is not only required to exponentially stretch the
magnetic field lines but it needs to also arrange them in a constructive way so that when they are brought  arbitrarily close
they can survive the effect of diffusion.  A quantitative measure of these effects can be obtained by 
multiplying Eq.(\ref{dynamo}) by ${\bf b}$, space averaging and dividing by $\langle{\bf b^2}\rangle$ and finally 
using Eq.(\ref{str_eig}) to obtain
\begin{equation}\label{ener}
\gamma +\partial_t \ln\left[ \langle \tilde{\bf b}^2 \rangle \right]  = \frac{ \langle{\bf b \cdot (\nabla u) b}\rangle}{\bf\langle b^2 \rangle} - \eta \frac{ \langle (\nabla {\bf b } )^2 \rangle }{\langle \bf b^2 \rangle}.
\end{equation}
Performing a time average and using the fact that ${\bf \tilde{b}}$ is bounded the last equation can be written as
\begin{equation}\label{gamma}
\gamma = \gamma_s - \gamma_d
\end{equation}
where $\gamma_s$ is the time average of the first term in the right hand side of Eq.(\ref{ener}) 
                                                                             and expresses the injection rate of energy by stretching,
while $\gamma_d$ is the time average of the second term                      and expresses the dissipation rate.
A constructive flow then has large $\gamma_s$ and small $\gamma_d$.
This 
can be obtained by the stretch-twist-fold mechanism \cite{STF} that aligns the stretched magnetic field lines
so that they have the same orientation. 
It is expected to be achieved most efficiently if the flow is helical.

Helicity is the other ingredient that is expected to improve dynamo action.
It is a measure of the lack of reflection symmetry of the flow \cite{Moffatt1978} and is related to the
linking number of the flow lines. Although in general it is not necessary \cite{Hughes1996}, it has been thought to improve
dynamo action and it is required for $\alpha^2$ dynamos \cite{Parker1955,Braginsky1964,Steenbeck1966}.
It is also considered important for the the generation of the large scale magnetic fields
that are observed in the universe \cite{Meneguzzi1981,Brandenburg2001,Vishniac2001,Alexakis2006}.

In this work we are going to restrict ourselves in a family of flows that is 
both fully helical and are known to have chaotic flow lines, namely the ABC flows.
The ABC flows include a wide range of expected dynamo behaviors that covers 2D flows, 
slow and fast dynamos. Particular members of this  family have been well studied
for dynamo action and this allows for a comparison with previous results. 
This choice is rather restrictive since it is not known a priori if the optimal dynamo flow 
belongs in the family of ABC flows.
However, they provide a tractable set of flows to examine and a good starting guess.
  
The ABC flows are reviewed in detailed the next section.

\section{The ABC family}

The ABC flow is named after V. Arnold \cite{Arnold1965}, E. Beltrami \cite{Beltrami1889}, and S. Childress \cite{Childress1970}, 
and is explicitly given by:
\begin{eqnarray}
u_x = A \sin(k_u z) + B \cos(k_u y), \nonumber  \\
u_y = C \sin(k_u x) + A \cos(k_u z),            \\
u_z = B \sin(k_u y) + C \cos(k_u x). \nonumber
\end{eqnarray}
It is an incompressible periodic flow with 4 independent parameters $A,B,C$ and $k_u$.
The flow has the property:
\begin{equation}\label{belt}
{\bf w} =\nabla \times {\bf u } = k_u {\bf u} 
\end{equation}
for all values of $A,B,C,k_u$.
As a result it is an exact solution of the Euler equations. It has been studied both for its properties 
as a solution of the Euler equation, its relation to chaos \cite{Arnold1983,Dombre1986,Podvigina1994} and for dynamo instability 
\cite{Galloway1986,Galanti1992,Archontis2003} but only for limited values of the parameters. Here it is attempted to uncover the 
dynamo properties for the whole family.

With no loss of generality we can restrict ourselves only to flows of fixed wavenumber $k_u$ and fixed velocity amplitude 
$U=\sqrt{A^2+ B^2+ C^2}$. With this restriction and property \ref{belt}  
the energy of the flow $E=\frac{1}{2}U^2$, the enstrophy of the flow
 \beq \Omega = \frac{1}{2} \langle {\bf w\cdot w} \rangle = 
\frac{1}{2} k_u^2U^2, \eeq
and the helicity $H$ of the flow
 \beq H = \frac{1}{2} \langle {\bf w\cdot u} \rangle =
\frac{1}{2} k_uU^2   \eeq
have a fixed value.

For fixed kinetic energy the parameters $A,B,C$ live on the surface of a sphere of radius $U$ and 
can be parameterized using the spherical coordinates $\psi,\phi$:
\begin{eqnarray}
   A&=&U \cos(\psi), \\ 
   B&=&U \sin(\psi)\cos(\phi), \\ 
   C&=&U \sin(\psi)\sin(\phi).
\end{eqnarray}
Using the symmetries of the ABC flow (see \cite{Dombre1986}) we can restrict the examined parameter space.
%
The flow is invariant under the transformations 
\begin{eqnarray}
                   & [A,z] & \to   [-A,z-\pi/k_u], \\ 
                   & [B,y] & \to   [-B,y-\pi/k_u], \\  
\mathrm{and}\quad  & [C,x] & \to   [-C,x-\pi/k_u].
\end{eqnarray}
These symmetries allow to restrict the investigation to only positive values of A,B and C
and thus reduce the examined parameter space to the range $[0,\pi/2]$ for both angles $\phi$ and $\psi$.
Since there is no preferred direction between $(x,y,z)$ the growth rate is also going to be
independent under permutations, e.g. $(B,C) \to (C,B)$. More precisely the flow is invariant under the transformations

\begin{equation}\label{S1}
\left[ \begin{array}{c}
                     (A,B,C) \\
                     (x,y,z) 
       \end{array}
\right]
\to
\left[ \begin{array}{c}
                     (A,C,B) \\
                     (\frac{3\pi}{2k_u}-y,\frac{3\pi}{2k_u}-x, \frac{3\pi}{2k_u}-z) \\
       \end{array}
\right]
\end{equation}
\begin{equation}\label{S2}
\left[ \begin{array}{c}
                     (A,B,C) \\
                     (x,y,z) 
       \end{array}
\right],
\to
\left[ \begin{array}{c}
                     (B,A,C) \\
                     (\frac{3\pi}{2k_u}-x,\frac{3\pi}{2k_u}-z, \frac{3\pi}{2k_u}-y) \\
       \end{array}
\right]
\end{equation}
\begin{equation}\label{S3}
\left[ \begin{array}{c}
                     (A,B,C) \\
                     (x,y,z) 
       \end{array}
\right],
\to
\left[ \begin{array}{c}
                     (C,B,A) \\
                     (\frac{3\pi}{2k_u}-z,\frac{3\pi}{2k_u}-y, \frac{3\pi}{2k_u}-x) \\
       \end{array}
\right].
\end{equation}
These symmetries allow to inter-change the values of any of the three parameters $A,B,C$
and were used  to improve the estimates of the growth rate from the numerical simulations. 
Finally, the change $k\to -k$ changes the sign of the helicity of the flow.
This change however does not alter the resulting growth rate of the dynamo. Thus only positive values of $k$ are considered.

Depending on the values of the parameters $A,B,C$ the flow can have eight stagnation points where all three components of 
the velocity are zero. These points exist only if the square of each the parameters $A,B,C$ is smaller than the sum of 
the square of the other two \cite{Dombre1986} ({\it ie} their squares can form a triangle). The importance of the existence or 
absence of stagnation points was  emphasized in \cite{Archontis2007}. It was noticed that 
the developed magnetic structures changed from from ``cigar-shaped" in the presence of stagnation points to 
``ribbon-shaped" in their absence.  

\begin{figure}
\includegraphics[width=8cm]{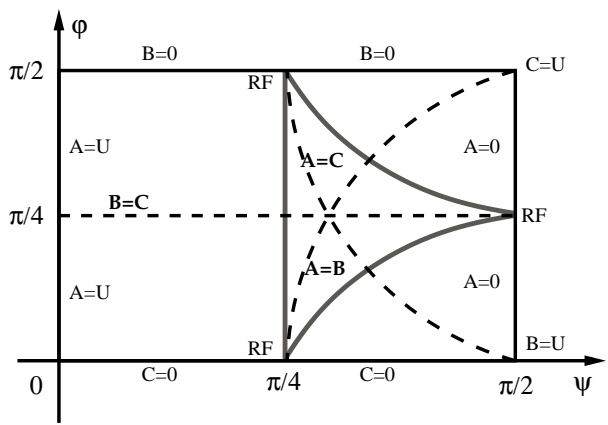}
\caption{\label{d1} A sketch of the parameter space for the ABC family of flows.
RF marks the location of the Roberts Flow. The dashed lines indicate the location
where two of the three parameters ($A,B,C$) are equal and at their intersection is the 1:1:1 flow. 
The grey lines enclose the region where stagnation points exist.}
\end{figure}

Figure \ref{d1} demonstrates the examined parameter space in the spherical coordinates ($\psi,\phi$).
Some of the points in this graph represent flows of special significance that are described in what follows.
For ($\psi=\pi/2$, $\phi=\pi/2$), ($\psi=\pi/2$, $\phi=0$) and  ($\psi=0$) two of the three parameters $A,B,C$ are zero
    ($A=B=0$,                                 $A=C=0$          and     $B=C=0$   respectably).
The flow corresponding to these points is 2D and thus there is no dynamo, $\gamma<0$.
 
For ($\psi=\pi/2$) we have A=0, for ($\phi=\pi/2$) B=0, and for ($\phi=0$)  C=0; for these values of ($\psi,\phi$) for which
one of the three parameters $A,B,C$ is zero, the resulting flow is a $2\frac{1}{2}$D flow and thus a slow dynamo.
The Roberts flow is a special flow in this subset for which the two nonzero parameters are equal.
It has been studied for slow dynamo action in \cite{Roberts1970,Roberts1972,Soward1987}.
It corresponds to the values ($\psi=\pi/2,\phi=\pi/4$),  ($\psi=\pi/4,\phi=\pi/2$), ($\psi=\pi/4,\phi=0$) and
in figure \ref{d1} it is marked as RF.

Flows which have two of the three parameters equal have additional symmetries and as will be shown in the result 
section they are important.
These flows are located along
                    the line $\phi=\pi/4$ for B=C,
                    the line $\psi=\mathrm{\arctan}(1/\cos(\phi))$     for A=B 
             and    the line $\psi=\mathrm{\arctan}(1/\sin(\phi))$     for A=C.
These lines are shown by dashed lines in figure \ref{d1} and divide the space in six compartments.
Each of these compartments is equivalent to the others due to the symmetries in Eq.(\ref{S1},\ref{S2},\ref{S3}). 
Thus each of these compartments will have the same number of maxima and minima of the growth rate.

When all three parameters are equal A=B=C the flow has the largest number of symmetries.
This flow is the most studied one in the literature and it is going to be referred to as the 1:1:1 flow.
It is obtained for $(\phi=\pi/4$, $\psi=\mathrm{arctan}(\sqrt{2}))$ and is located in the intersection of the dashed 
lines in the diagram. 

Finally the region of the parameter space for which the ABC flow has stagnation points is enclosed by the 
grey lines in figure \ref{d1}.

\begin{figure}
\centerline{
\includegraphics[width=8cm]{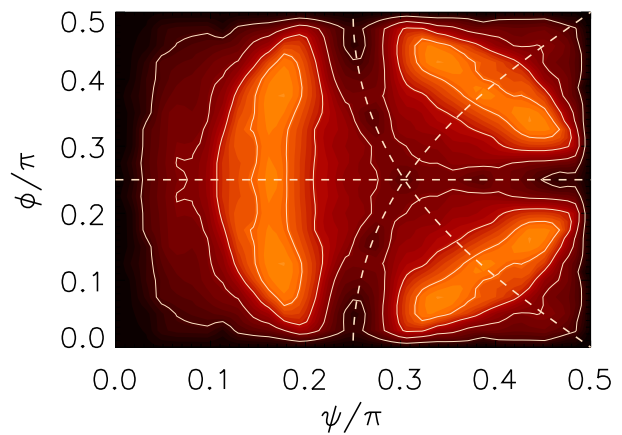}}
\caption{\label{L1} Color-scale plot of the finite time Lyaponov exponent $\lambda_\tau$ in the $\psi,\phi$ plane.
                    Bright colors imply large values of the exponents, while black imply zero or
                    close to zero values. The contour lines correspond to the levels $\lambda_\tau$=
                    0.02, 0.04, 0.06 and 0.08. The time of integration was $\tau=2\cdot 10^4$.}
\end{figure}
\begin{figure}
\centerline{
\includegraphics[width=8cm]{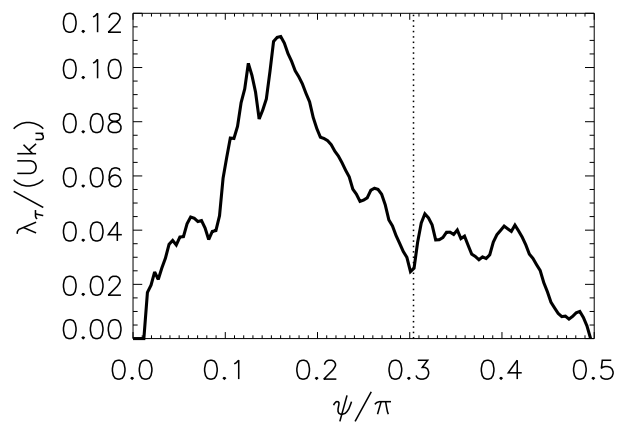}}
\caption{\label{L2} Plot of the finite time Lyaponov exponent $\lambda_\tau$ for $\phi=\pi/4$.
                    The time of integration was $\tau=10^6$.}
\end{figure}
ABC flows are known to be chaotic \cite{Arnold1983,Dombre1986,Galanti1992,Galloway1986,Podvigina1994}.
Finite time Lyaponov exponents provide a measure of chaos \cite{OTT}. 
The finite time Lyaponov exponent $\lambda_\tau(x_0)$ for a point $x_0$ is defined as:
\beq
\lambda_\tau(x_0) =\frac{1}{\tau} \ln\left[ \frac{|\delta {\bf x(\tau)}|}{|\delta {\bf x(0)}|} \right], 
\eeq
where $|\delta {\bf x(\tau)}|$ is the distance of two particles that at time $\tau=0$ they were placed 
infinitesimally close to $x_0$. 
If the flow is not ergodic, not all initial points $x_0$ of a chaotic flow lead to $\lambda_\tau>0$.
To measure thus $\lambda_\tau$ of the flow, an ensemble of initial points $x_0$ needs to be considered
out of which only those that belong to the chaotic subset will lead to $\lambda_\tau>0$.
Here, $\lambda_\tau(x_0)$ was calculated  for the ABC flows and for 8000 initial positions $x_0$ distributed uniformly
in the domain $[0,2\pi)^3$. The distribution function of the measured Lyaponov exponents was constructed and $\lambda_\tau$ of
the chaotic subset was determined as the location of the peek in the distribution function.

In figure \ref{L1} a color-scale plot of $\lambda_\tau$ is shown for the ($\psi,\phi$) plane and in figure 
\ref{L2} the finite time Lyaponov exponents are shown for $\phi=\pi/4$. 
It is worth noting that the $\lambda_\tau$ of the most symmetric flow 1:1:1 is a local minimum 
(see also \cite{Galanti1992}), while the largest values of  $\lambda_\tau$ appear 
                                        for $(\phi=     \pi/4$,   $\psi\simeq 0.155\pi)$  and 
                                        for $(\phi\simeq0.12\pi$, $\psi\simeq0.16\pi)$    and the equivalent points by symmetry. 

\section{Dynamo Results}

The advection diffusion equation (\ref{dynamo})  was solved in a triple periodic
domain of size $L=2\pi$ using a standard pseudo-spectral method and a third order in time
Runge-Kuta \cite{Minini_code1,Minini_code2}. The resolution used varied from $32^3$ grid points for small values of $R_{_M} (\lesssim 20)$ 
                                                                      up to $256^3$             for the largest values, $R_{_M}\gtrsim 500$.
 Each run was evolved for sufficiently long time 
until a clear exponential increase of the magnetic energy was observed and the growth rate was
calculated by fitting. 

The last parameter that needs to be defined is the ratio of the box size $L$ over which the magnetic field is allowed
to evolve in, to the period of the velocity field $2\pi/k_u$. Due to the periodicity the product $k_u L$ can 
only be integer multiples of $2\pi$. Here we are going to examine two cases $k_uL=2\pi$ where the two lengths are equal, and 
$k_uL=4\pi$ where the magnetic field can evolve on a larger scale.

\subsection{ ABC, $k_uL=2\pi$}
\label{zz}
\begin{figure*}
\includegraphics[width=7.5cm]{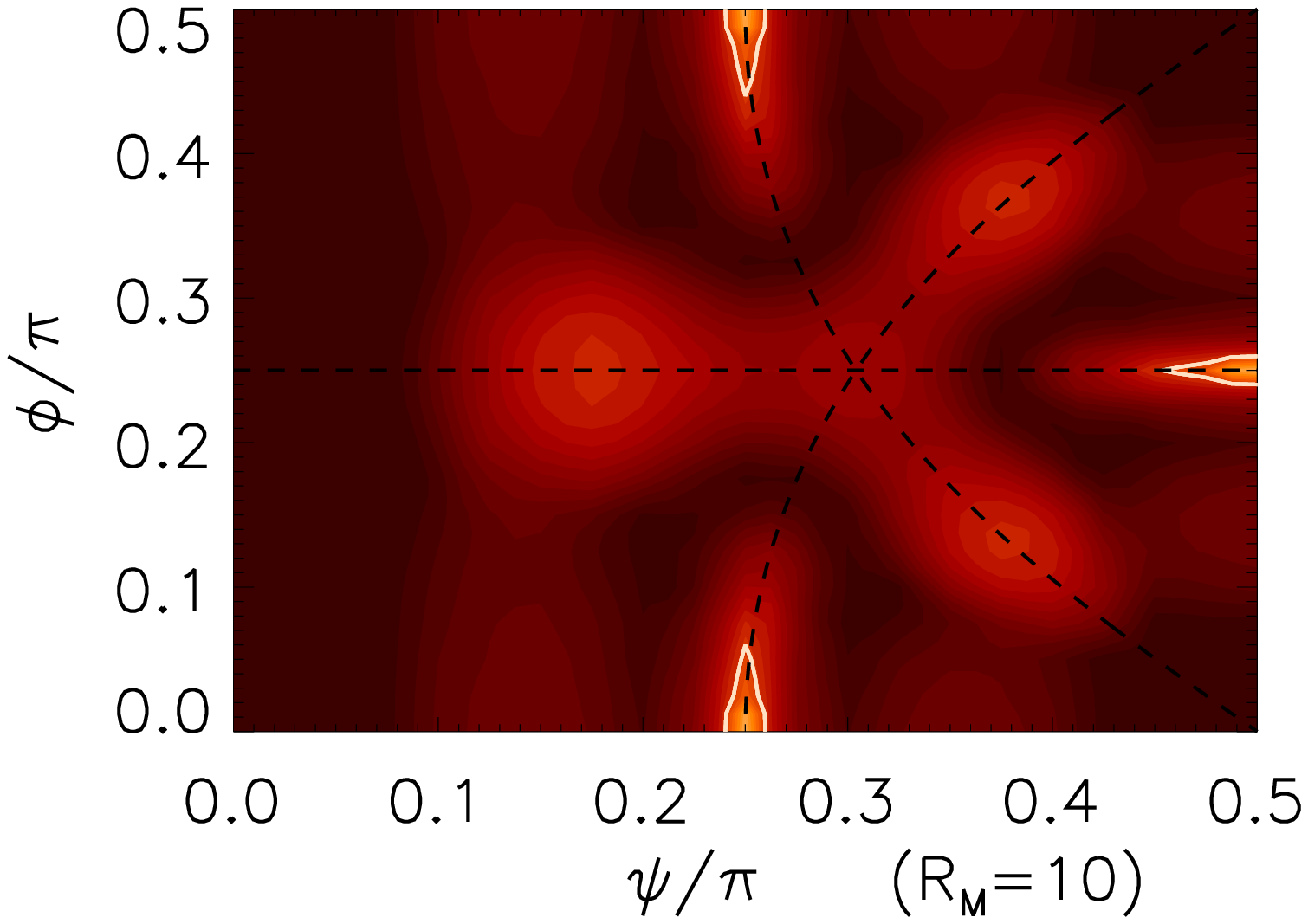} \hspace{0.5cm}
\includegraphics[width=7.5cm]{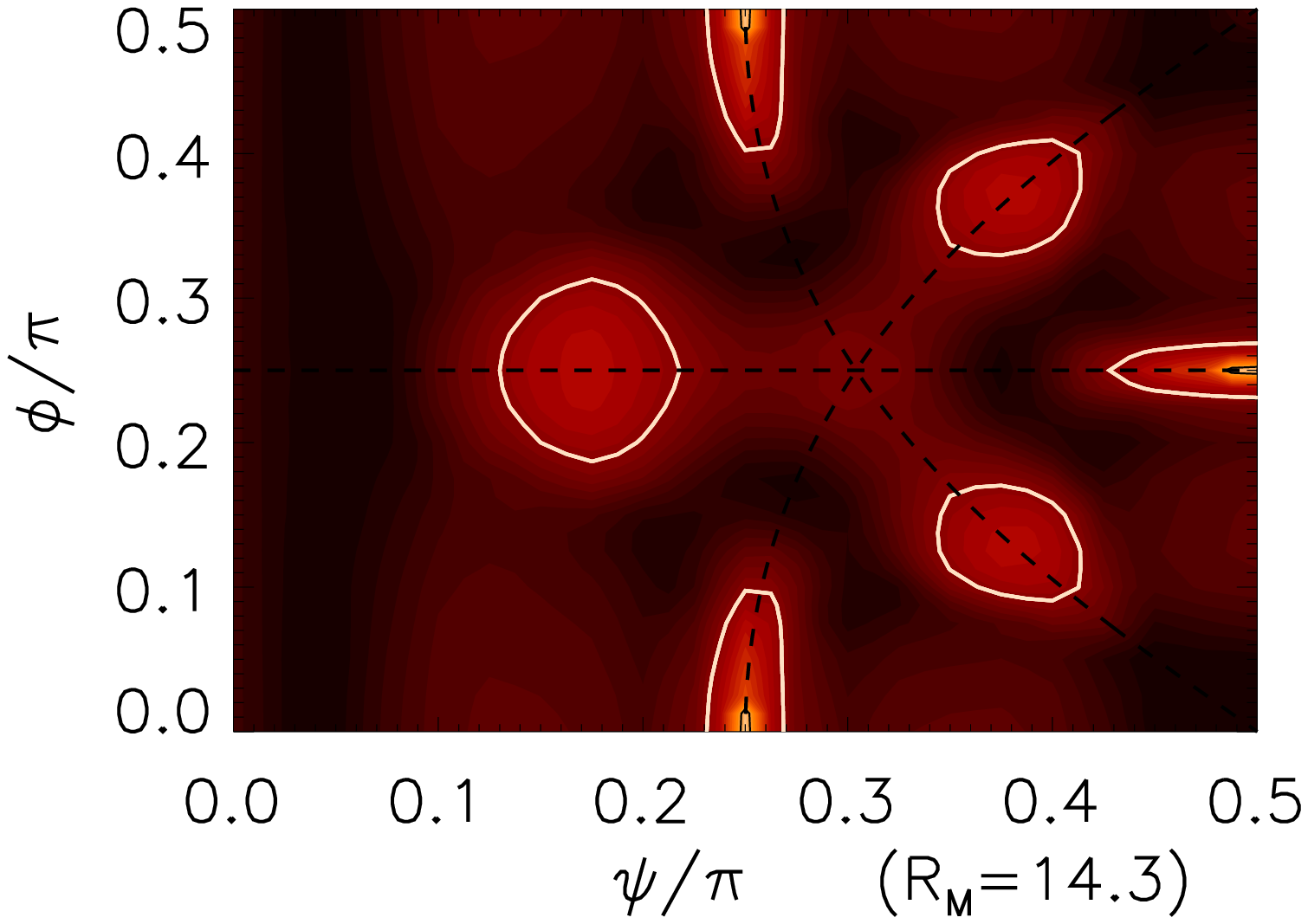} \\ \vspace{0.5cm}
\includegraphics[width=7.5cm]{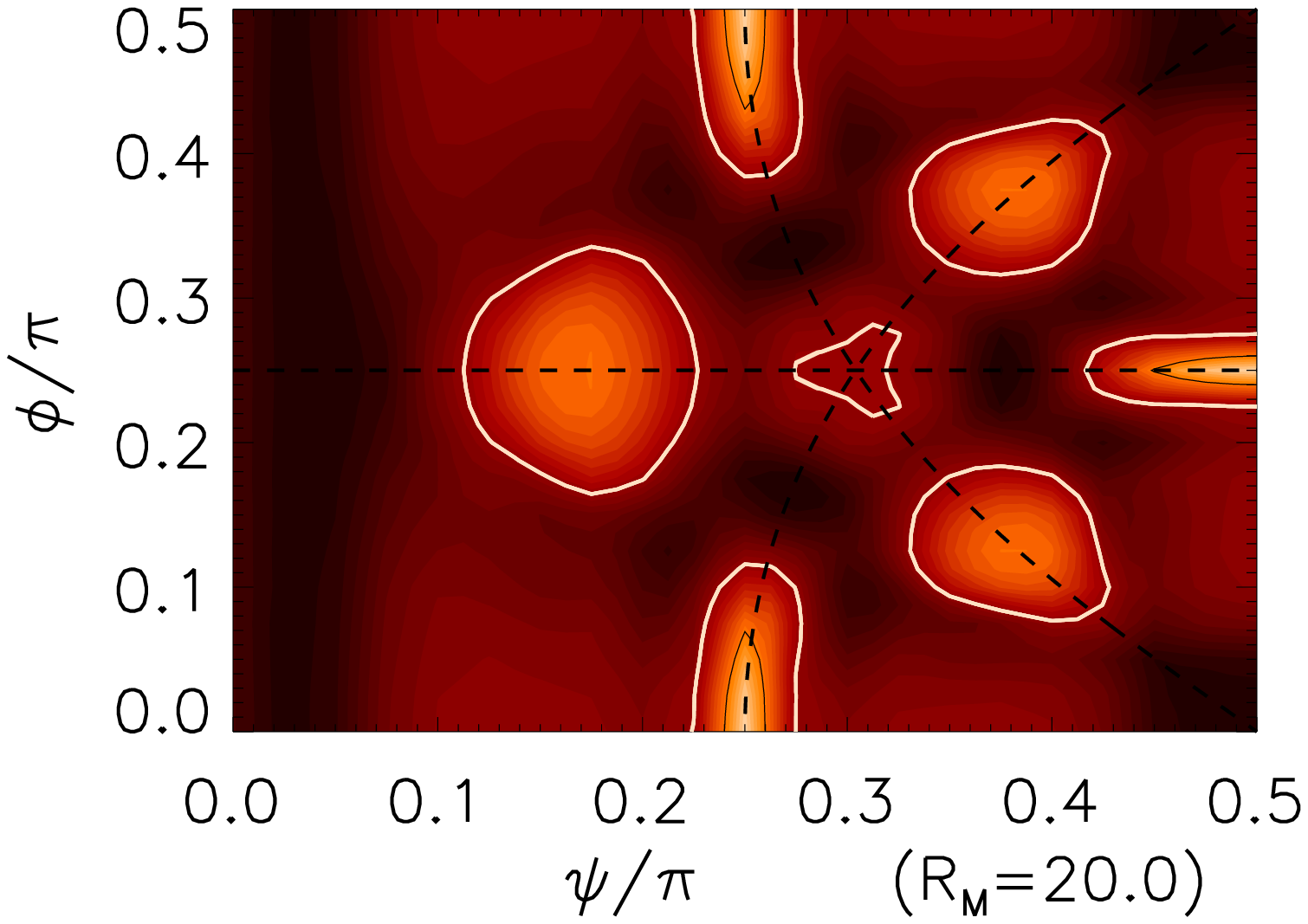} \hspace{0.5cm}
\includegraphics[width=7.5cm]{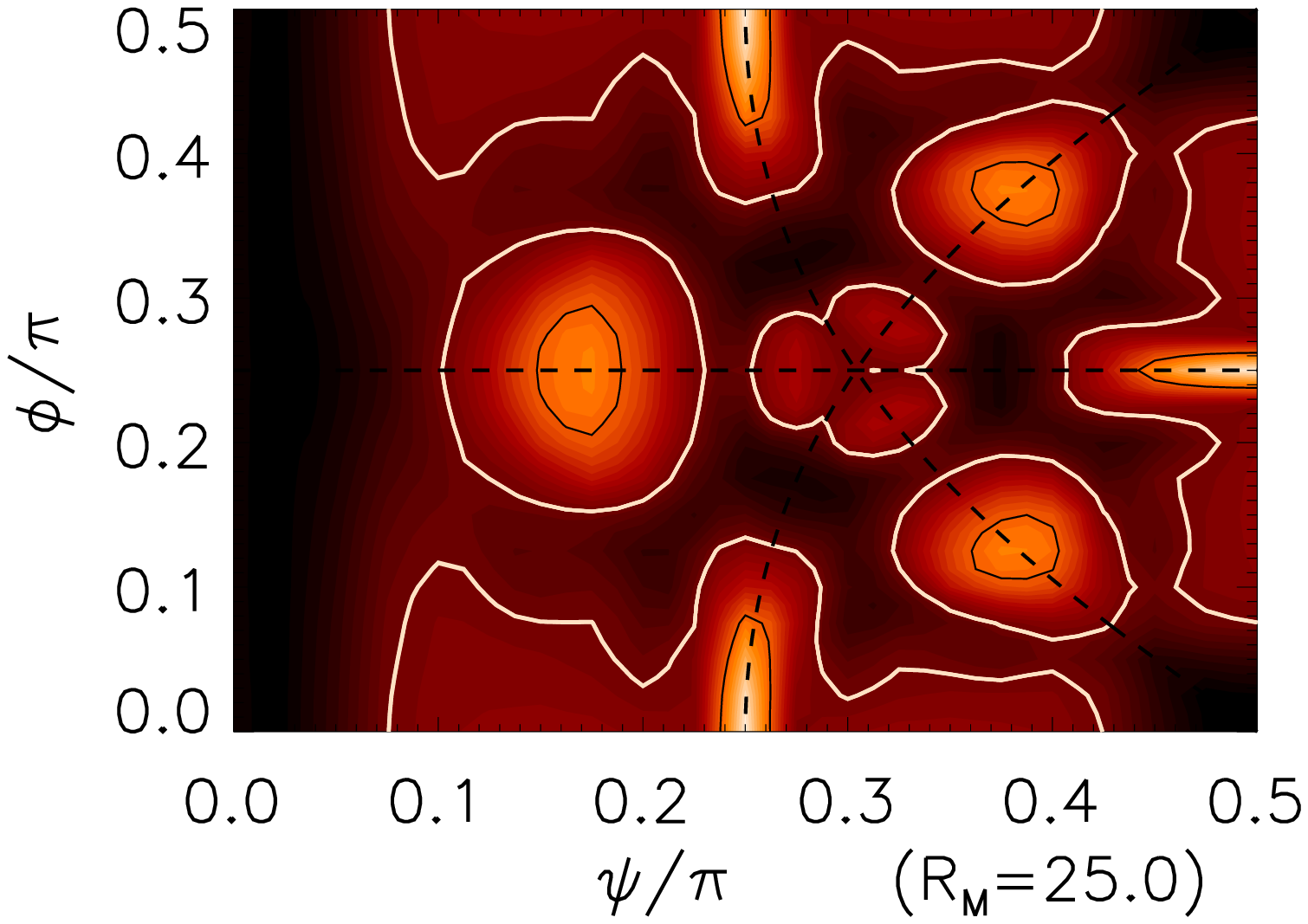} \\ \vspace{0.5cm}
\includegraphics[width=7.5cm]{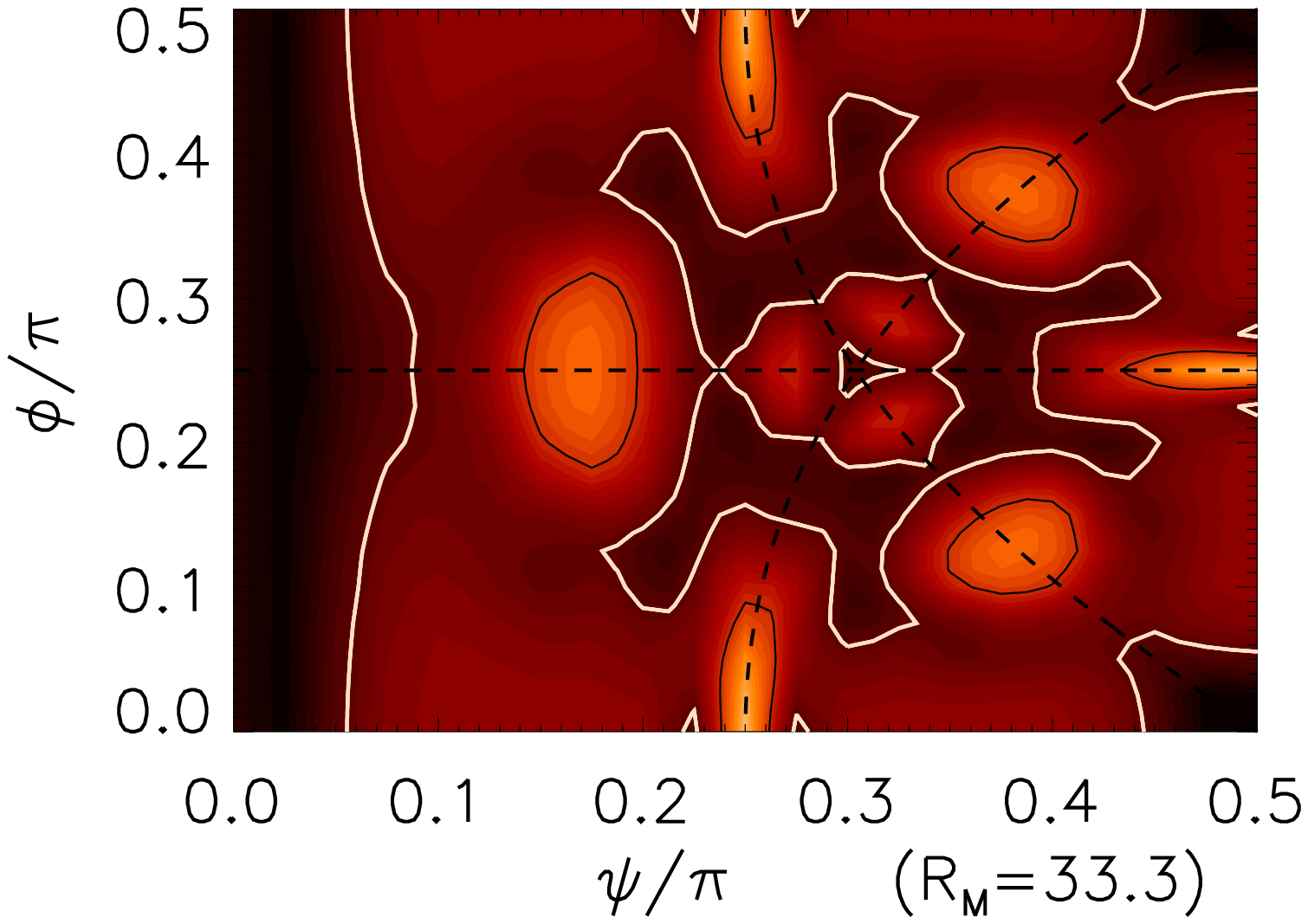} \hspace{0.5cm}
\includegraphics[width=7.5cm]{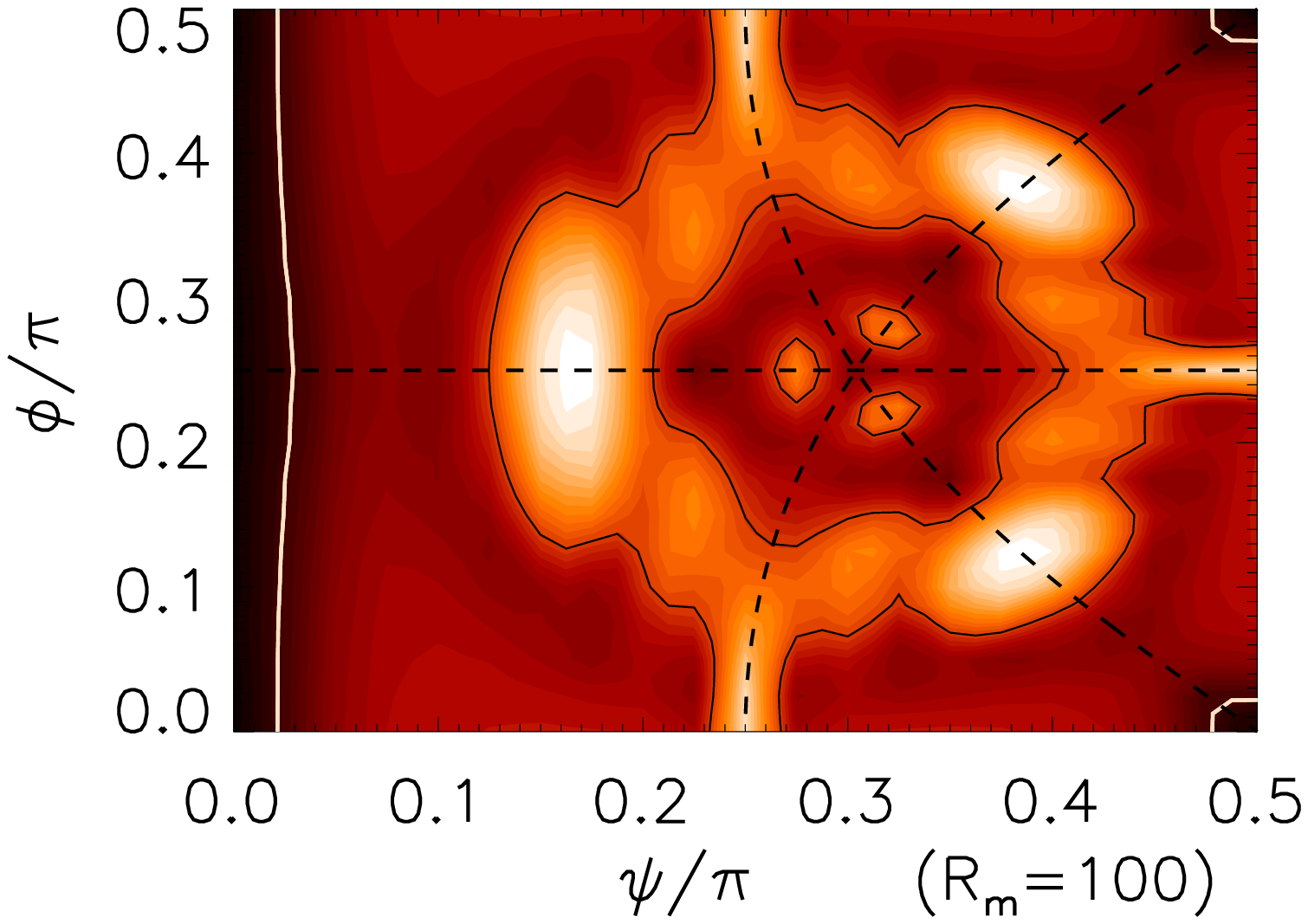}
\caption{\label{f2}  Color-scale images of the growth rate in the $\psi,\phi$ plane
                    for $k_uL=2\pi$ and for six different magnetic Reynolds numbers. Bright colors
                    indicate larger growth rate. The thick white lines are the contour lines
                    zero growth rate. Thin black lines are the contour lines
                    of growth rate $\gamma = 0.05Uk_u$. The dashed lines indicate the as in figure
                    \ref{d1} show the location for which to of the three parameters $A,B,C$ are equal. }
\end{figure*}

\begin{figure}
\centerline{
\includegraphics[width=8cm]{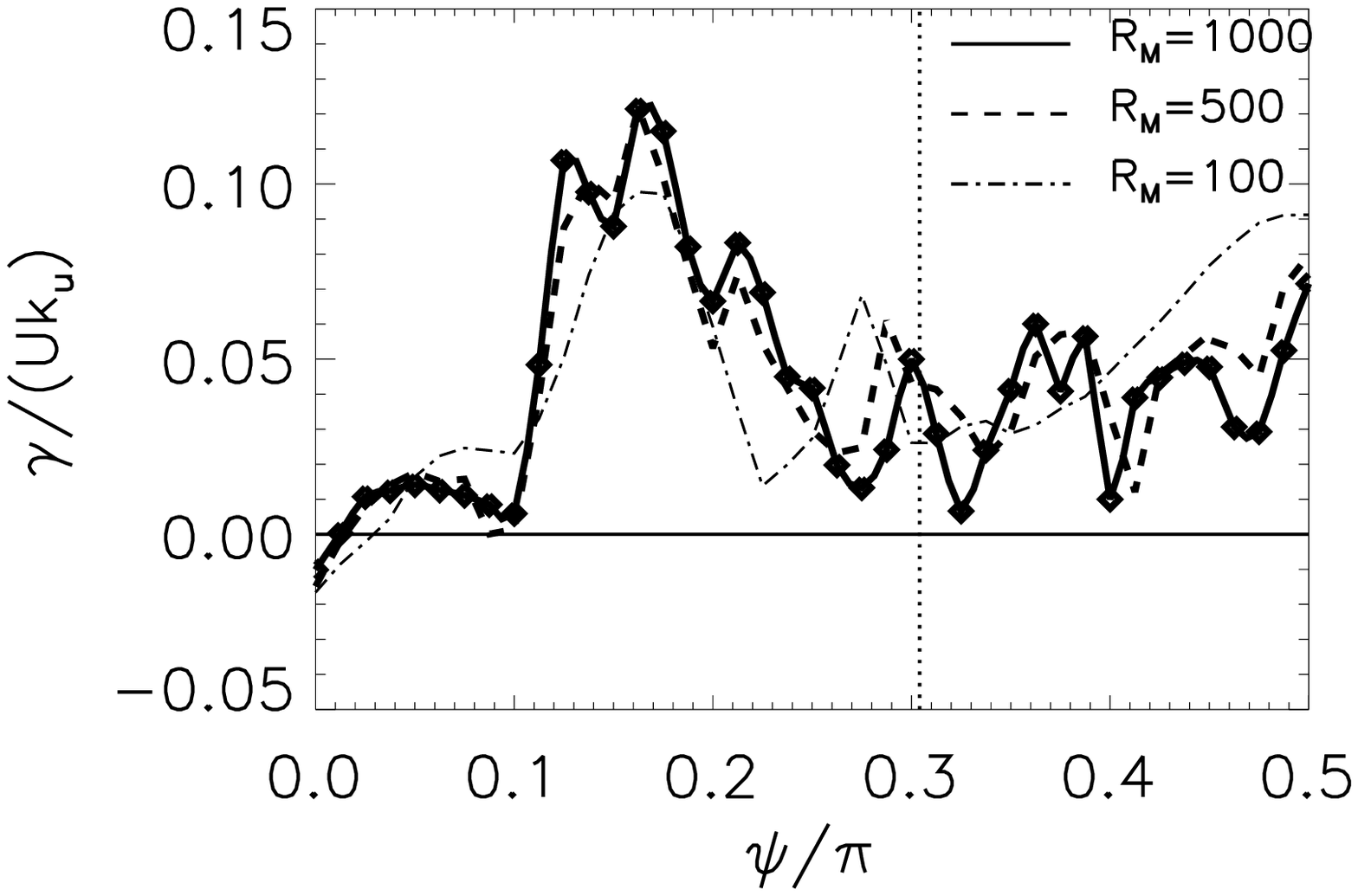}}
\caption{\label{f3} Non-dimensional growth-rate as a  function of $\psi$ for $\phi=\pi/4$ 
                    and for 3 different magnetic Reynolds numbers. The vertical dotted line marks the location
                    of the 1:1:1 flow.
                         }
\end{figure}

First the case $k_uL=2\pi$ is presented.
In figure \ref{f2} color-scale images of the measured growth rate are shown for six different values of 
$R_{_M}$.  Each figure corresponds to 200 different dynamo simulations. Using 20 different values of
$\psi$ in the range $[0,\pi/2]$ and 10 for $\phi$  in the range $[0,\pi/4]$. The symmetries in Eq.(\ref{S1}-\ref{S3}) 
were used to fill in the values of the growth rate on the whole domain and on a denser grid. 
In each panel bright colors correspond to larger growth rate. The thick 
white lines show the location of zero growth rate. The thin black lines indicate where the growth rate is 
$0.05Uk_u$. The dashed black lines as in figure \ref{d1} show the location on which two of the three parameters 
$A,B,C$ are equal. Finally, it is noted that the simulations in these runs were performed on $32^3$ and $64^3$ grid-point meshes.

As $R_m$ is increased the first flows that result in positive growth rates are the ones with two of the 
three parameters equal, while the third is equal to zero. This can be seen in the top 
left panel of figure \ref{f2} for $R_{_M}=10$  where most of the parameter domain has negative growth rate 
except the small bright regions at the end of the dashed lines. These flows correspond to a Roberts flow 
and are slow dynamos as discussed in the previous section. Thus, although they are slow, at small $R_{_M}$ 
they are the most efficient at producing a dynamo ($ie$ the fastest).

The next flows that become unstable are the flows for which two of the the three parameters are equal 
but smaller than the third. Thus they lie on the dashed lines in the graph ``opposite" the Roberts Flow. 
This can be seen in the right top panel of figure \ref{f2} that shows the growth rate for $R_{_M}=14.3$.
In terms of the angles they correspond to the values ($\psi \simeq 0.17\pi$, $\phi=\pi/4$), and 
                                                     ($\psi \simeq 0.38\pi$, $\phi=\pi/4\pm0.12\pi$). 
The exact location of these new maxima is shifting slowly away from the center as the magnetic Reynolds number is increased.
Note that this flow is very close to the flow for which the maxima for the
Lyaponov exponents in figures \ref{L1} and \ref{L2} were found. It is also close to the flow $A=5$, $B=C=2$ 
that was investigated  in detail in \cite{Archontis2007}, for this reason this flow is going to be referred as the 5:2:2 flow.
At this value of the Reynolds number the Roberts flow is still the  fastest dynamo in the family.

As the magnetic Reynolds number is increased further the most symmetric flow 1:1:1 also results in dynamo.
This is shown in the middle left panel of figure \ref{f2} for $R_{_M}=20$.  At this value of $R_{_M}$ the 
1:1:1 is a local maximum, but with smaller growth rate than the 5:2:2 flow and smaller than the 
Roberts flow that is still the fastest.

As $R_m$ is increased further the 1:1:1 flow  stops being a local maximum and transitions to a to a third-order 
saddle point (monkey saddle point). This can be seen in the middle right panel for which $R_{_M}=25$.
The local maximum of the 1:1:1 flow, that was present at $R_m=20$, splits to three local maxima that move along the dashed 
lines away from the 1:1:1 case who's growth rate has decreased. The growth rate for the 5:2:2 flow and the Roberts 
flow continues to increase. 

For $R_m=33.3$ (shown in the bottom left panel) the three local maxima that were initially located close to the 
1:1:1 flow have moved sufficiently away that the 1:1:1 flow  stops being a dynamo. This corresponds to the no-dynamo
window that was observed early on in \cite{Galloway1986}.

After further increase of $R_{_M}$ the  1:1:1 flow becomes a dynamo again (although not a local maximum anymore but still
a saddle point). For $R_{_M}=100$ shown in the bottom right panel most of parameter space is resulting in dynamo action,
with only exception the small areas close to the 2D flows ($\psi=\pi/2$, $\phi=\pi/2$), ($\psi=\pi/2$, $\phi=0$) and  ($\psi=0$). 
The growth rate of the Roberts flow has started to decrease and the fastest dynamo is given by the 5:2:2 flow.
At this value of  $R_{_M}$, the ``topography" of growth rate in the parameter space has become much more complex, with 
new local maxima appearing between the Roberts flow and the 5:2:2 flow. 
 
For larger values of $R_{_M}$ grids larger than $64^3$ are needed and it is computationally too expensive cover the 
whole parameter domain. 
Instead the investigation will be limited to flows which lie along the dashed lines where most of the maxima are located.
In figure \ref{f3} we show the growth rate for three different values of the magnetic Reynolds number with the smallest
value being equal to the value used in the last panel of figure \ref{f2}. The 5:2:2 flow results in the fastest dynamo
at the largest value of $R_{_M}=1000$. At this value of $R_{_M}$ the 5:2:2 peek has moved to $\psi \simeq 0.16\pi$.
Furthermore, for $R_{_M}>500$ two new local maxima appear close to the 5:2:2 flow for slightly
smaller and slightly larger values of $\psi$. 
The local maximum which at small values of $R_{_M}$ was located at the 1:1:1 flow
appears to return close to the 1:1:1 point and thus the most symmetric flow comes close to a local maximum again. 
Finally the slow decrease of the growth rate of the Roberts flow can be observed.

\begin{figure}
\centerline{
\includegraphics[width=8cm]{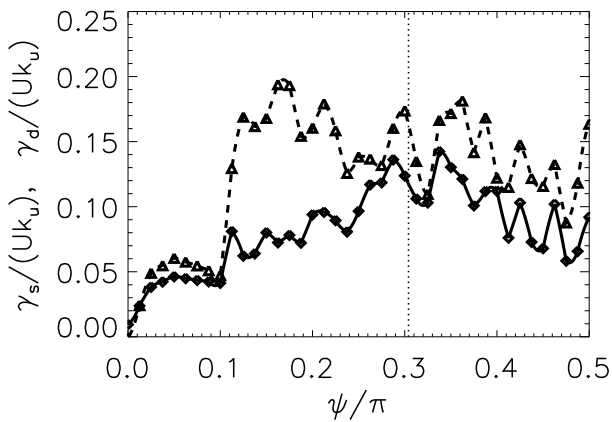}}
\caption{\label{sd1} The  rates $\gamma_s$ (dashed line, triangles) and $\gamma_d$ (solid line, diamonds)
defined in Eq. (\ref{gamma}) for the $k_uL=2\pi$ case and for $R_{_M}=1000$. The vertical dotted line marks the location
                    of the 1:1:1 flow.}
\end{figure}

In figure \ref{sd1} the two rates $\gamma_s$ (dashed line, triangles) and $\gamma_d$ (solid line, diamonds) 
defined in Eq.(\ref{gamma}) are shown for the largest examined $R_{_M}=1000$.
The difference between the two curves gives the growth rate. The ratio of the two curves shows the percentage 
of the injected energy that is dissipated. Thus a constructive flow 
(in the sense that it aligns magnetic field lines pointing in the same direction)
is expected to have a small  value of $\gamma_d$ compared to $\gamma_s$.
The flows close to the 5:2:2 flow ($\psi=0.16\pi$) that have the largest growth rates are more efficient not 
only due to the larger stretching rate $\gamma_s$ that does not vary a lot, but also due to the relatively small value of $\gamma_d$.
In the range $\psi=0.1\pi$ to $\psi=0.25\pi$ half of the energy injected by stretching goes to magnetic field amplification.
For values of $\psi$ out of this range, only a small fraction of the injected energy goes to field amplification  
while the rest is going to the small scales where it is dissipated.

Beyond the $\phi=\pi/4$ symmetry line other local maxima of the growth rate were detected although it was not feasible to cover
the entire parameter space. Here it is just mentioned that a local maximum was observed at ($\psi=0.2\pi,\phi=0.1\pi$)
with growth rate close to the 5:2:2 flow $\gamma\simeq 0.12$. 
    
\subsection{ ABC $kL=4\pi$}
\begin{figure*}
\includegraphics[width=7.5cm]{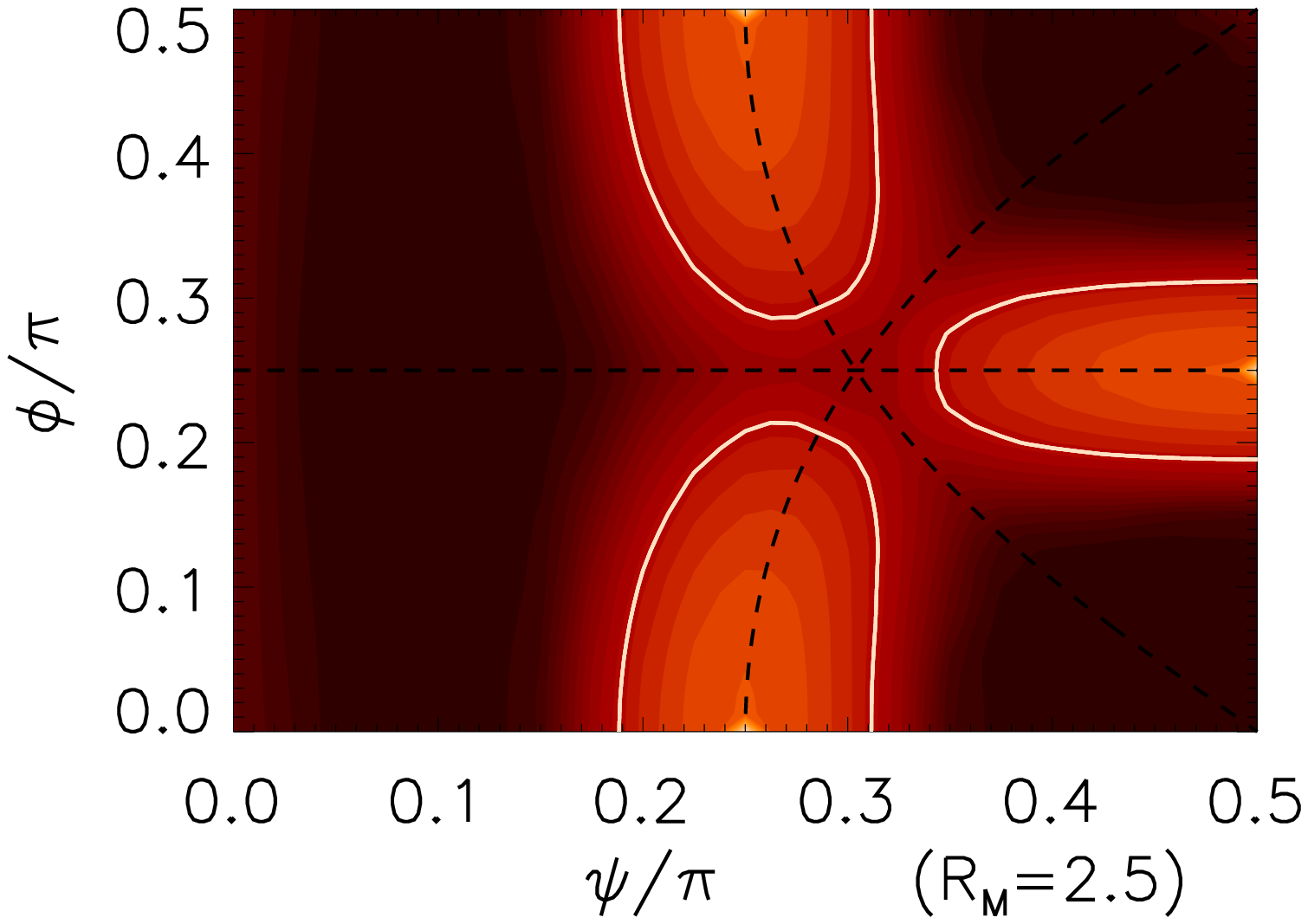} \hspace{0.5cm}
\includegraphics[width=7.5cm]{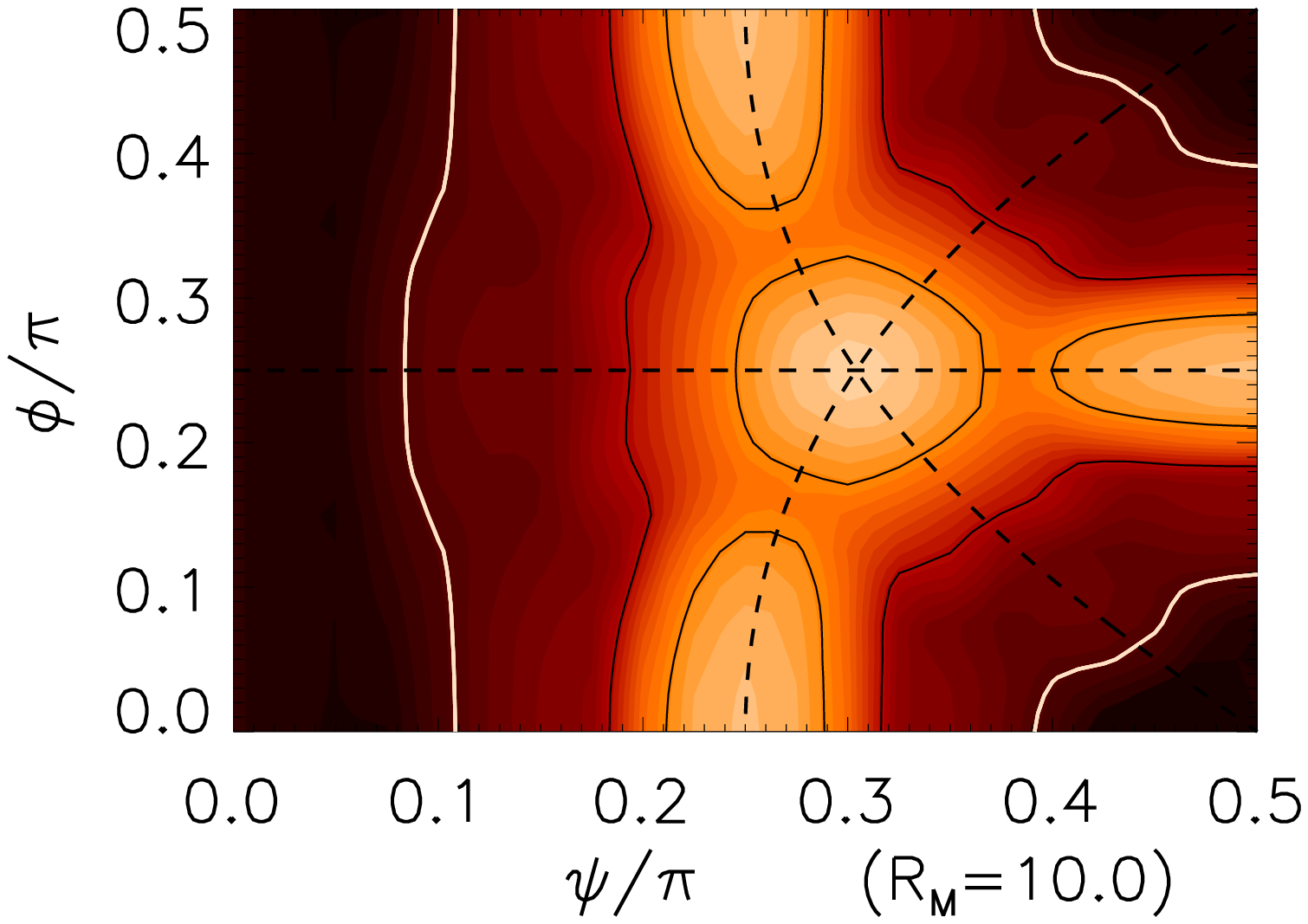} \\ \vspace{0.5cm}
\includegraphics[width=7.5cm]{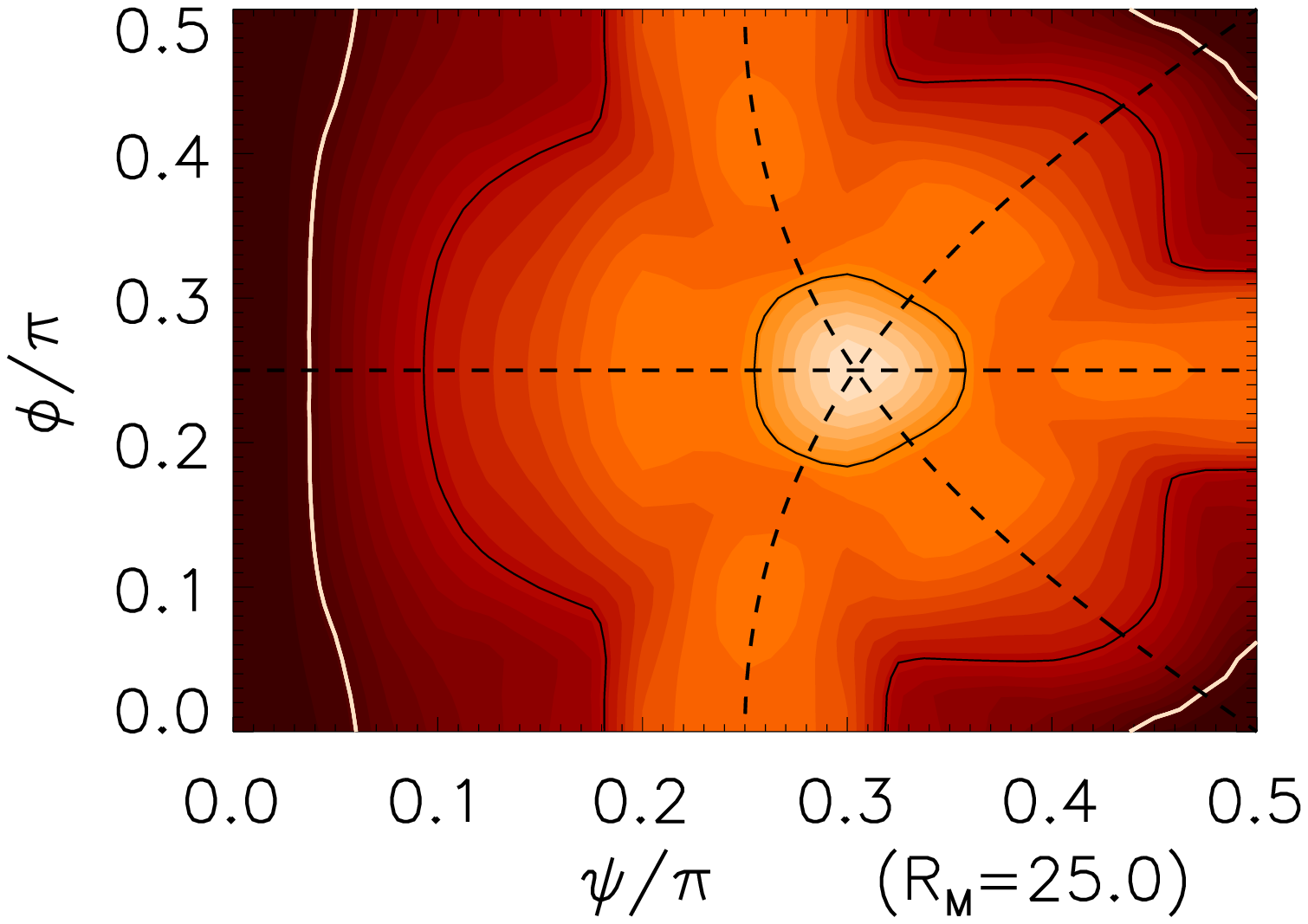} \hspace{0.5cm}
\includegraphics[width=7.5cm]{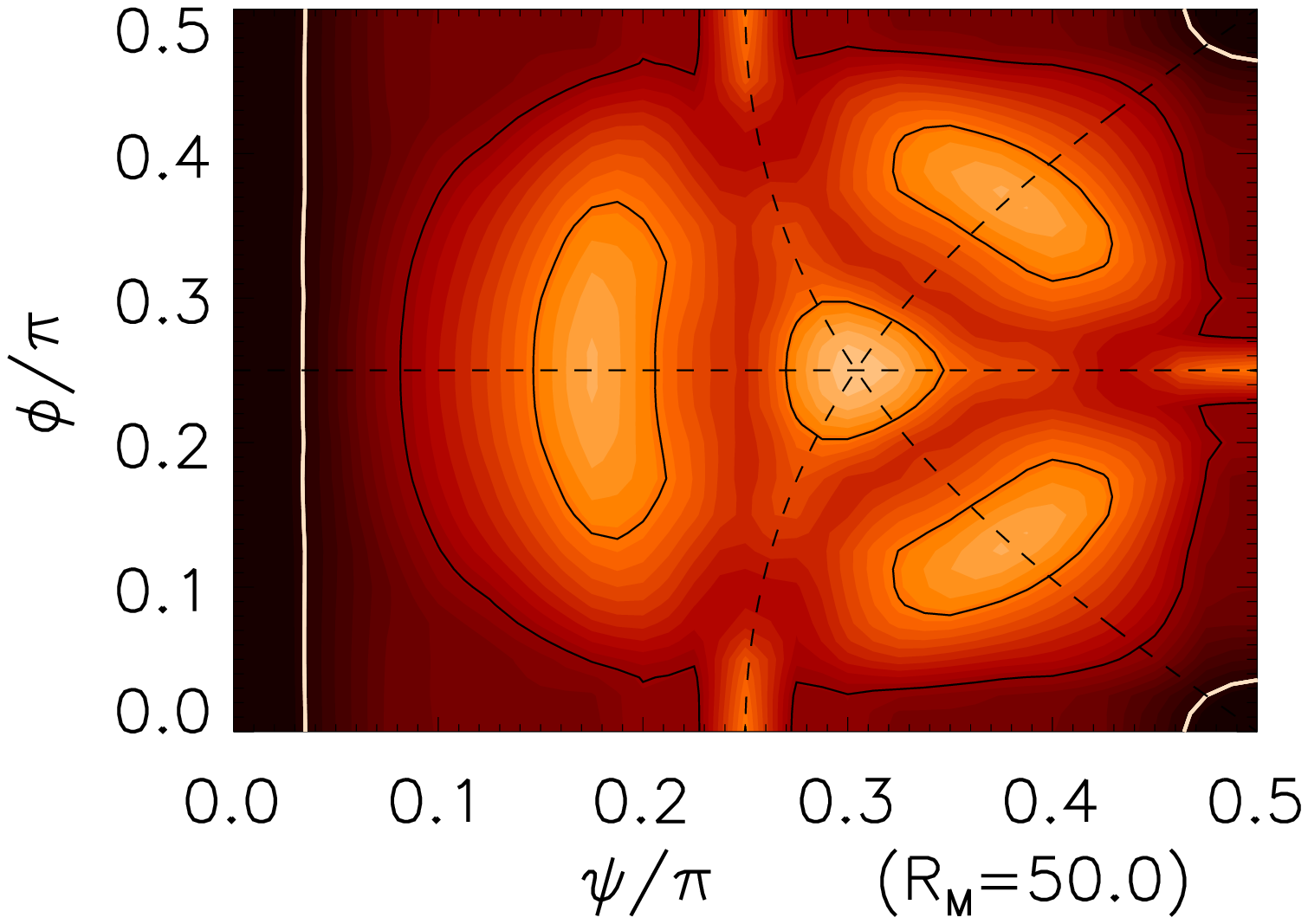}
\caption{\label{f4} 
                    Color-scale images of the growth rate in the $\psi,\phi$ plane
                    for $k_uL=4\pi$ and for four different magnetic Reynolds numbers. Bright colors
                    indicate larger growth rate. The thick white lines are the contour lines
                    zero growth rate. Thin black lines are the contour lines
                    of growth rate $\gamma = 0.05Uk_u$ and $\gamma = 0.1Uk_u$. The dashed lines indicate the as in figure
                    \ref{f2} show the location for which to of the three parameters $A,B,C$ are equal. }
\end{figure*}
\begin{figure}
\centerline{
\includegraphics[width=8cm]{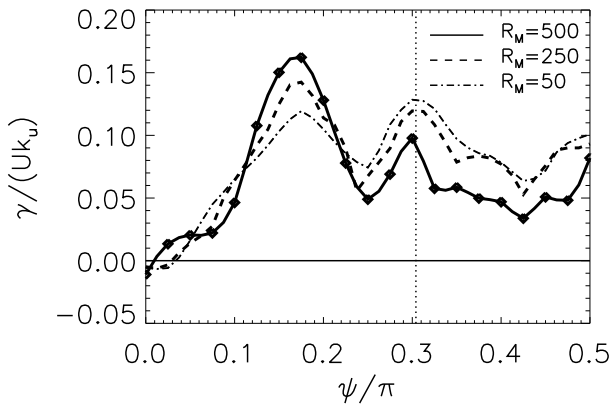}}
\caption{\label{f5} 
                    The growth rate as a function of $\psi$ and for $\phi=-\pi/4$
                    for $k_uL=4\pi$ and three different Reynolds numbers.The vertical dotted line marks the location
                    of the 1:1:1 flow.}
\end{figure}

The case $k_uL=4\pi$ is examined in this section. Despite the fact that the flow is the same as in the $k_uL=2\pi$ case, the
results are different  due to the additional space in which the magnetic field is allowed to evolve. The extra space gives rise 
to new modes that can develop with different growth rates. Although the modes of the $k_uL=2\pi$ case are still present and grow 
at the same rate, they are not necessarily the fastest. Considering that in a numerical simulation only the fastest growing 
mode is observed, in the $k_uL=4\pi$ case the observed mode will then be at least as fast as the $k_uL=2\pi$ case.

As before, the first flows that result in dynamo 
are the slow dynamos of the Roberts flow for which two of the three parameters $A,B,C$ are equal and the third is zero.
This case is shown in the right panel of figure \ref{f4} for $R_{_M}=2.5$. Note that in this case the dynamo instability 
appears at much smaller values of $R_{_M}$. It is also remarkable that the mode whose growth rate peaks for the Roberts flow, 
appears to continuously extend all the way to the 1:1:1 flow that is a saddle point at this stage.  
 
As $R_{_M}$ increases further, the 1:1:1 flow becomes a dynamo whose growth rate is a local maximum in the ($\psi,\phi$) plane.
This is shown in the top right panel of figure \ref{f4} that corresponds to $R_m=10$. Note also that this is contrary to
the $k_uL=2\pi$ case for which the 5:2:2 flow was the second flow to result in dynamo. In the $k_uL=4\pi$ case and for
this value of $R_{_M}$ there is no observed local maximum close to the 5:2:2 flow.
  
As $R_{_M}$ is further increased the growth rate of the 1:1:1 flow is increased. At
$R_{_M}=25$ the 1:1:1 flow exceeds the Roberts flow in growth rate and it is the fastest dynamo for all ABC flows. 
This can be seen in the bottom left panel of figure \ref{f4}. This is somehow surprising since this flow was never the 
fastest in the $k_uL=2\pi$ case.

At even larger $R_{_M}$ however the growth rate of the 1:1:1 ceases to increase while the 5:2:2 becomes a local maximum
and obtains comparable values with the 1:1:1 flow. 
This is shown in the bottom right panel in figure \ref{f4}. 

The growth rate for larger values of $R_{_M}$ was calculated only along the symmetry line $\phi=\pi/4$.
It is shown as a function of $\psi$ and for three different values of $R_{_M}$ in figure \ref{f5}.
The smallest value of $R_{_M}$ corresponds to the results of the bottom right panel of figure \ref{f4}. 
As the magnetic Reynolds number is increased the growth rate of the 1:1:1 flow is decreasing while at the 
same time the growth rate of the 5:2:2 flow is increasing. At the largest examined value of $R_{_M}$ the 
fastest dynamo is given by the 5:2:2 flow with a growth rate $\gamma/(k_uU)=0.16$ which is larger than its 
growth rate in the $k_uL=2\pi$ case. 
\begin{figure}
\centerline{
\includegraphics[width=8cm]{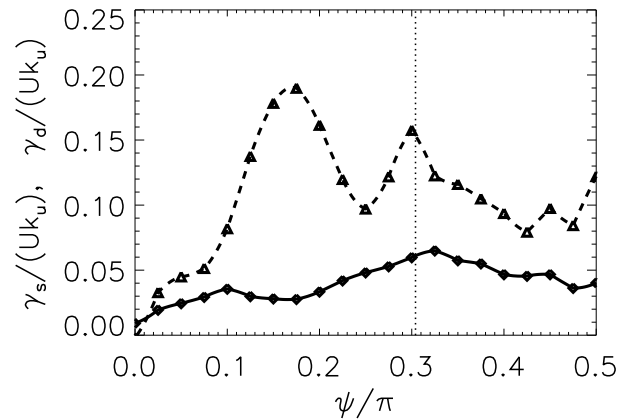}}
\caption{\label{sd2} The  rates $\gamma_s$ (dashed line, triangles) and $\gamma_d$ (solid line, diamonds)
defined in Eq. (\ref{gamma}) for the $k_uL=4\pi$ case and for $R_{_M}=500$.The vertical dotted line marks the location
                    of the 1:1:1 flow.}
\end{figure}

As in the previous section we plot in figure \ref{sd2} the two growth rates $\gamma_s$ (dashed line, triangles)
and $\gamma_d$ (solid line, diamonds) for $R_{_M}=500$. In this case the 
the ability of the 5:2:2 flow to align field lines reducing dissipation 
is even more pronounced. Only one fifth of the injected energy is cascading to the dissipated scales while the rest
is going in the amplification of the magnetic field. Unlike the $k_uL=2\pi$ case the 1:1:1 flow is also being more
``constructive" with less than half of the energy going to dissipation. 

It is also worth comparing the general behavior of the growth rate with the results in the previous section.
Although the fastest dynamo flows appear at the same location their growth rates are different thus it is not the
same dynamo modes that are observed in the two cases. 
Also, in the $k_uL=4\pi$ case the dependence of the growth rate on the flow  at the large $R_{_M}$ is less complex than in the
$k_uL=2\pi$ case with less local maxima and a smoother in general behavior.
These differences indicate that the box size plays an important role in the dynamo behavior.

\section{Summary and Conclusions}
 
In this work the entire family of ABC flows was examined for dynamo  action. The dynamo growth rate
was calculated as a function of the magnetic Reynolds number $R_{_M}$ and for two length scales $k_uL=2\pi$ and $k_uL=4\pi$.
The questions that this work was attempting to answer were: 
(i)    which flow has the smallest critical magnetic Reynolds number $R_{_{MC}}$,
(ii)   given $R_{_M}$ which flow has the largest growth rate $\gamma/(Uk_u)$, and 
(iii)  which flow leads to the largest growth rate in the limit $R_{_M}\to\infty$. 
Although these questions can be posed for a larger family of flows the ABC-flows constitute a first step
in obtaining some understanding.

For this restrictive perhaps family of flows the answer to the first question is a simple one:
The Roberts flow results in dynamo for the smallest value of $R_{_M}$.
This is true for both the  $k_uL=2\pi$ and the  $k_uL=4\pi$  case.
Thus in small Reynolds numbers a well organized flow can do much better than a rapidly stretching (chaotic) flow.

At larger Reynolds numbers new dynamo modes became unstable and a
number of bifurcations are observed that lead to a complex ``topography" of the growth rate.  
As $R_{_M}$ was increased this complexity is further increased and more local maxima appeared. 
This is particularly true for the $k_uL=2\pi$ case, while for the $k_uL=4\pi$ case a 
smoother behavior was observed. 

Inspecting the growth-rate for a large number of flows as is done in this work also gives a wider perspective on the
dependence of magnetic eigenmodes of the flow on $R_{_M}$. 
Some of the observed dynamo modes of a given flow can be be related (by continuous transform) to the modes of different flows. 
For example for small values of $R_{_M}$ the slowest decaying mode of the 1:1:1 flow is related to the dynamo mode 
of the Roberts flow (see figure \ref{f4} top left panel).
Thus the various bifurcations that can be observed by looking the growth rate of a single flow, 
can be interpreted as shifting or enlargement of local maxima in this wider point of view. 
The no-dynamo window of the 1:1:1 flow is such an example, which is the outcome of splitting and shifting of the 
initial maximum at the 1:1:1 point.

Finally, for relative large $R_{_M}$ the 5:2:2 flow 
($\psi \simeq 0.16\pi$, $\phi=\pi/4$) has the fastest growing mode (from the examined flows) in both cases 
($k_uL=2\pi$ and $k_uL=4\pi$). 
However, if this continues to be true for even larger values of the magnetic Reynolds number cannot be concluded from the present data.
$R_{_M}=1000$ is still far from the $R_{_M}\to\infty$ limit as can be seen from the finite value of the growth rate
of the Roberts flow which is a slow dynamo.
Furthermore, as noted at the end of section (\ref{zz}), flows that were not on the $\phi=\pi/4$ symmetry line were found  
with growth rates similar to the 5:2:2 flow.
The increased complexity  of the growth rate 
as $R_{_M}$ increases makes it harder to estimate the fastest dynamo flow. 
If this continues, then the location of the fastest flow in the ($\psi,\phi$) plane might not converge to a single point 
in the limit $ {R_{_M}\to\infty}$ and question (iii) might not even have an answer.

On the other hand the Lyaponov exponents, whose value does not depend on $R_{_M}$ do show some clear maxima,
which gives hope that a fastest dynamo flow in the $R_{_M}\to\infty$   limit  exists.
However, although a correlation of the growth rate with the Lyaponov exponents is observed, it is definitely 
not sufficient to explain the dependence of the observed growth rates, at least not at the examined values of 
$R_{_M}$. In particular it is  observed that the flow with the largest growth rate is close to the flow with 
the largest Lyaponov exponent. Nevertheless, the general dependence of the growth rate and of the Lyaponov exponent on the 
flow is quite different, with local maxima appearing at different locations. 

In addition it was found that the 5:2:2 flow that lead to the fastest growing mode besides having large stretching rate it
was also very efficient at organizing the magnetic field lines as to minimize the magnetic energy dissipation.
Furthermore at the examined values of $R_{_M}$ the $k_uL=2\pi$ and the  $k_uL=4\pi$  cases showed significant differences
although the magnetic field lines are advected by the same flows. Thus the growth rate can not be determined by the stretching
statistics of the flow alone. If these differences cease to exist at larger $R_{_M}$ is a question of future work.

Besides investigating larger $R_{_M}$ there many other obvious extensions of this work. 
First it would be interesting to extend these results in a larger family of flows, that also include non-helical flows.
Harmonic velocity fields could be such a generalization.

A differently oriented approach would consider turbulent dynamos. In this case, instead  of prescribing the flow,
a body force would be prescribed and the flow would be allowed to evolve dynamically. 
In such a study different limits of the  kinetic Reynolds number $Re$ would lead to different results. 
In the limit $1\ll Re \ll R_{_M}$ dynamo growth rates depend on the small velocity scales and are possibly universal. 
In the other limit $1 \ll R_{_M} \ll Re$ it has been shown that the large scale flow plays an important role especially for
$R_{_M}$ near its threshold value \cite{Ponty2005,Scheko2005,Ponty2011}.

Finally the properties of dynamos beyond the linear regime, where our understanding is much more limited,
is also a problem of considerable interest. 
At the nonlinear stage, both the saturation levels of the magnetic energy and the involved length scales 
(large or small scale dynamo) depend strongly on the large scale properties of the flow.
Thus a systematic study of a large number of flows can be helpful in that respect.

These issues are going to be pursued in the authors future work.

%
\begin{acknowledgments}
Computations were carried out on the CEMAG computing center at LRA/ENS and on the 
CINES computing center, and their support is greatly acknowledged.
\end{acknowledgments}


%


\end{document}